# The influence of adatom diffusion on the formation of skyrmion lattice in sub-monolayer Fe on Ir(111)

Mingming Shuai[1,2], Yulong Yang[1,2], Haiming Huang[1,2], Rui Song[1,2], Yi Zhu[1,2], Yanghui Liao[1,2], Yinyan Zhu[1,2,4], Xiaodong Zhou[1,2,4]*, Lifeng Yin[1,2,3,4,5]*, and Jian Shen[1,2,3,4,5]*

[1] State Key Laboratory of Surface Physics, Institute for Nanoelectronic Devices and Quantum Computing, and Department of Physics, Fudan University, Shanghai 200433, China.
[2] Shanghai Qi Zhi Institute, Shanghai 200232, China
[3] Shanghai Research Center for Quantum Sciences, Shanghai 201315, China.
[4] Zhangjiang Fudan International Innovation Center, Fudan University, Shanghai 201210, China.
[5] Collaborative Innovation Center of Advanced Microstructures, Nanjing 210093, China.

*Emails: zhouxd@fudan.edu.cn, lifengyin@fudan.edu.cn, shenj5494@fudan.edu.cn

## Abstract

Room temperature grown Fe monolayer (ML) on the Ir(111) single crystal substrate has attracted great research interests as nano-skyrmion lattice can form under proper growth conditions. The formation of the nanoscale skyrmion, however, appears to be greatly affected by the diffusion length of the Fe adatoms on the Ir(111) surface. We made this observation by employing spin-polarized scanning tunneling microscopy to study skyrmion formation upon systematically changing the impurity density on the substrate surface prior to Fe deposition. Since the substrate surface impurities serve as pinning centers for Fe adatoms, the eventual size and shape of the Fe islands exhibit a direct correlation with the impurity density, which in turn determines whether skyrmion can be formed. Our observation indicates that skyrmion only forms when the impurity density is below $0.006/nm^2$, i.e., 12 nm averaged spacing between the neighboring defects. We verify the significance of Fe diffusion length by growing Fe on clean Ir(111) substrate at low temperature of 30 K, where no skyrmion was observed to form. Our findings signify the importance of diffusion of Fe atoms on the Ir(111) substrate, which affects the size, shape and lattice perfection of the Fe islands and thus the formation of skyrmion lattice.

## 1. Introduction

Magnetic skyrmions are non-collinear spin textures characterized by a topological winding number[1]. They are induced by the competition of collinear Heisenberg exchange between atomic spins and non-collinear Dzyaloshinskii–Moriya (DM) interaction in non-centrosymmetric magnetic compounds[2] or in thin films with broken inversion symmetry[3]. Skyrmions behave as particles whose creation, movement and annihilation form the basis of their applications in information storage and logic devices[4-6]. Fe monolayer (ML) grown on Ir(111) single crystal surface attracts particular research interests as it hosts a stable nano-skyrmion lattice at zero magnetic field, which is ideal for skyrmion's applications[7-14]. Density functional theory calculations suggest that the formation of skyrmion lattice in the system results from the interplay between Heisenberg exchange, DM interaction and a higher order four-spin interactions[15]. These magnetic interactions are ultimately related to the Fe atomic structure on the Ir(111) surface. For example, while a square skyrmion lattice is routinely observed in fcc-stacked Fe ML on Ir(111), a hexagonal skyrmion lattice exists in hcp-stacked Fe ML on Ir(111)[11]. The correspondence between the stacking order of Fe atom on Ir(111) substrate and the symmetry of the formed skyrmion lattice indicates that the existence of skyrmion lattice itself should also be susceptible to the perturbations to the spatial arrangement of Fe atoms on Ir(111) surface.

The spatial arrangement of adatoms on a substrate surface often depends sensitively on the adatom diffusion, which can be controlled by substrate defects or substrate temperature. In this work, we investigate how the Fe adatom diffusion affects the skyrmion lattice formation of Fe ML on Ir(111) substrate using spin-polarized scanning tunneling microscopy (SP-STM)[16]. The substrate defects are impurity atoms segregated from the bulk internal to the surface, whose density can be controlled by sputtering and annealing procedures upon substrate cleaning process. We observe that high (low) defect density gives rise to fractal (triangular) shaped Fe islands and the absence (appearance) of skyrmion lattice, indicating that Fe adatom diffusion plays an important role in determining the spatial arrangement of Fe atoms and thus the formation of the skyrmion lattice. This is further confirmed by our observation that Fe deposition on nearly defect-free Ir(111) at low temperatures also results in fractal shaped islands without the presence of skyrmion lattice. The apparent correlation between Fe adatom diffusion and the formation of skyrmion lattice underlies the interplay between structural, electronic and magnetic degrees of freedom in the system[17].

## 2. Experimental details

A sub-monolayer of Fe was thermally deposited on an Ir(111) single crystal surface. Prior to the Fe deposition, the Ir(111) single crystal was treated by cycles of Ar ion sputtering and subsequent annealing at 1300 K in an ultrahigh vacuum system with a base pressure of $7 \times 10^{-9}$ mbar. During the annealing, impurity atoms will segregate from the bulk internal to the surface forming pinning centers to prevent adatom diffusion. These impurities are mostly Rh, C, Pt and W for the Ir single crystal used in our experiment. To get rid of the impurity atoms, the Ir(111) substrate was further annealed in oxygen atmosphere ($2 \times 10^{-6}$ mbar) at a lower temperature (~1200 K) for an extended time. It was found that the impurity density depends on the annealing time in oxygen allowing one to have a controllable impurity density on the Ir(111) surface. After the substrate treatment, Fe was thermally deposited at a rate ~0.08Å/min onto the Ir(111) substrate from a rod heated by e-beam bombardment. The substrate was at room temperature during the Fe deposition. The deposition rate was monitored by a quartz microbalance calibrated by STM morphology. SP-STM measurements were carried using a 0.25 mm diameter Fe tip at 5 K under a 2000 Oe magnetic field perpendicular to the substrate surface, which ensures that the magnetic moment at the tip apex points along the field direction and has a sensitivity to the out-of-plane component of the sample magnetization.

## 3. Results and discussion

Fig. 1(a-e) show a large area topography of bare Ir(111) substrate surface with varying surface impurity densities from 0.4212/nm$^2$ to 0.0024/nm$^2$ prepared by aforementioned procedures. The change of the impurity density has a dramatic effect on the Fe growth. Fig. 1(f-j) show the topography of typical Fe islands grown on Ir(111) substrate with the corresponding surface impurity density level displayed on its top. While Fe islands are generally monolayer high, their shape and size depend strongly on the surface impurity density. For surface with the highest impurity level (Fig. 1(a)), Fe islands are small with an irregular shape indicating a limited surface diffusion of Fe adatoms (Fig. 1(f)). As the impurity level is lowered, the Fe islands become larger indicating an improved surface diffusion of Fe adatoms. The shape of the islands is fractal-like, which is caused by the suppressed diffusion of adatoms at the island edges (Fig. 1(g-h)). In the clean limit with very few surface impurities, the Fe islands are triangular-shaped with large size, indicating that the Fe adatoms have an easy diffusion path both on the Ir(111) surface and along the island edges with the absence of the pinning centers (Fig. 1(i-j)). To quantify the correlation between the impurity density level and the shape of the Fe island, we adopt a geometry complex transform (GCT) method to determine the similarity of a specific Fe island shape to a perfect equilateral triangle. A quantity termed "similarity factor" is obtained from GCT. The smaller the factor is, the closer to an equilateral triangle an Fe island shape is. We calculate an averaged similarity factor of Fe monolayer islands seen in each image of Fig. 1(f-j) and plot it as a function of the surface impurity density in Fig. 1(k). A monotonic change is observed, i.e., the shape of an Fe island gradually deviates from a perfect equilateral triangle with an increasing surface impurity density. The sensitive dependence of the island size and shape on the impurity density reflects the important role of adatom diffusion upon the non-equilibrium growth at room temperature[18,19]. It has been known that triangular shaped islands tend to be formed as a result of the three-fold rotational symmetry of the underlying Ir(111) substrate[8]. However, a significant amount of surface impurities obstructs such kind of Fe island formation by suppressing the Fe adatom diffusion on the surface via pinning effect. As a result, Fe adatoms form islands with either irregular or fractal shapes. Triangular Fe islands are only formed

when enhanced surface diffusion is allowed on nearly clean surface.

The change of Fe island shape has a direct consequence on the formation of skyrmion lattice. Fig. 2 shows high resolution STM topography images and its height differential of Fe islands formed on Ir(111) surface with highest impurity density in (a-b) and lowest impurity density in (c-d), which corresponds to the situation in Fig.1(a) and (e), respectively. Note that height differential images are presented here for better visualization of the height change. Fe islands in Fig. 2(c-d) show a clear square lattice signal contrast, whose magnetic origin can be confirmed by performing the STM measurement under opposite magnetic fields as shown in Fig. 2(e-h). In the SP-STM experiment, the spin-polarized tunneling current depends on the relative orientation between the tip and the local sample magnetization, giving rise to a signal contrast with a magnetic origin in the constant-current STM topographic image. Fig. 2(e-f) are two constant-current STM topographic images of the square lattice (presumably skyrmion) collected under the +0.25 T and -0.25 T magnetic field, respectively. The applied magnetic field is big enough to flip the magnetic direction of the ferromagnetic Fe tip but not enough to change the skyrmion spin structure. As expected, the signal contrast of such a square lattice is reversed between Fig. 2(e) and (f), confirming its magnetic origin. To highlight this signal reversal, Fig. 2(g-h) show the sum and the difference of Fig. 2(e) and (f), respectively. The difference ((e)-(f)) will maximize the magnetic contrast, while the sum ((e)+(f)) highlights the features with a non-magnetic origin. Apparently, Fig. 2(h) shows a bigger signal contrast than that of Fig. 2(g). Therefore, such a square lattice is identified as magnetic nano-skyrmion[7,8,15]. Instead of a skyrmion lattice, a maze structure is observed in the Fe islands in Fig. 2(a-b). One also sees Fe double-layer regions in Fig. 2(a-b) as inferred from the line-cut height profile (inset of Fig. 2(a)). This double layer region shows no skyrmion, consistent with previous studies[8,13].

It is also interesting to look at the situation on Ir(111) surface with an intermediate impurity density level. Fig. 3(a) shows a large area topography of Fe islands grown on Ir(111) substrate with an impurity density corresponding to Fig. 1(d). The height differential image is presented in Fig. 3(b). There are five Fe islands in the area. Interestingly, while two fcc-stacked islands (island 1&2) on the top left show square skyrmion lattice with flat surfaces, the other three hcp-stacked islands (island 3, 4&5) with an opposite orientation on the right show no skyrmion lattice. Moreover, subtle differences exist among the hcp-stacked islands: Island 5 shows a mixture of the "bright" and the "dark" regions with a difference in the apparent height, while islands 3 and 4 display a spatially uniform "bright" state. It is known that the hexagonal skyrmion lattice is more difficult to resolve in hcp-stacked Fe islands than the fcc-stacked ones which may correspond to the situation of island 3 and 4[11]. The apparent mixed-phase state in island 5 indicates that there exist other non-skyrmion spin states competing with the skyrmion state in the Fe/Ir(111) system. As mentioned before, the magnetic ground state of Fe/Ir(111) will be determined by the delicate balance between different competing magnetic interactions. The differences of skyrmion formation in these islands indicate that at intermediate impurity level, the diffusivity of the Fe adatoms is at a critical value where statistical fluctuation of impurity distribution may affect the Fe atom spatial arrangements and thus the magnetic interactions. As a consequence, skyrmion formation varies in different Fe islands.

We also perform a field dependent STM measurement on such a non-skyrmion phase to obtain more insights of its signal contrast. Fig. 4(a) shows an example of such non-skyrmion phase in an Fe island. Similar to the islands 3 and 4 in Fig. 3, it looks very smooth in the STM topographic image. However, no skyrmion lattice is observed (Fig. 4(b)). Fig. 4(c) and (d) are two zoomed-in STM images of a square area in the monolayer region denoted in Fig. 4(a). They are taken under -0.25 T and +0.25 T magnetic field, respectively. One can see fine features in such a zoomed-in image, such as the peanut shape protrusion. Moreover, the signal contrast doesn't change very much between Fig. 4(c) and (d). Fig. 4(e) and (f) are the sum and the difference of Fig. 4(c) and (d). We choose two characteristic locations to compare its line-cut profile between Fig. 4(e) and (f). For the peanut shape protrusion, one gets a larger signal contrast in the sum image than that of the difference image (Fig. 4(g)), indicating its structural origin. However, it becomes less clear for the flat terrace region as shown in the Fig. 4(h). Both the sum and the difference images show a small and comparable signal contrast, indicating a mixture of the structural and the magnetic origin for this region. We must note that, such a comparison between the tip-up and the tip-down measurement also suffers from a certain ambiguity. For example, a non-fully spin polarized STM tip can reduce the magnetic contrast of a pure magnetic spin state. On the other hand, if the magnetization of both the STM tip and the sample's spin state flip with the field, one gets the same signal contrast when reversing the field. Further investigations are needed to identify the magnetic state of such a non-skyrmion phase.

The nano-skyrmion lattice is argued to be a magnetic ground state in Fe/Ir(111) system in previous work[15]. Our experiment shows that the impurity density of the Ir(111) surface plays a crucial role in achieving such a magnetic ground state. Such cleanness is directly tied to the Fe adatom diffusion on Ir(111) substrate surface, which is the most important kinetic process in thin film growth. A smooth and uniform film can only be formed with sufficient adatom mobility. In the extreme case of zero mobility, an adatom will stay where it lands, yielding a very rough growth front[20]. For epitaxial growth of metal on close-packed surfaces like the Fe/Ir(111) system, previous experiments showed both irregular-shaped and triangular-shaped islands depending on the detailed film/substrate combination. Kinetic Monte-Carlo simulations are performed to rationalize the correlation between these observations and the adatom diffusion along the edge and around the corner[21]. Our experimental observation further correlates the Fe island morphology with the skyrmion formation. As discussed above, the formation of skyrmion depends on the energy competition between different magnetic interactions which, in turn, relies on the spatial arrangement of Fe atoms on the Ir(111) surface. For example, the Fe-Ir hybridization should be involved to mediate the magnetic interaction between Fe atomic spins. Such hybridization sensitively relies on the orbital overlap of Fe and Ir atom and their relative spatial position. The irregular or fractal shapes of the Fe islands on Ir(111) imply that the diffusion of Fe adatoms is limited, which likely results in a less perfect lattice structure of the Fe islands. This constitutes a big perturbation to the skyrmion magnetic ground state. Therefore, it is not surprising to see the absence of skyrmion lattice on irregular-shaped and fractal-shaped Fe islands.

A controlled experiment was conducted to validate our hypothesis. Fe was grown on a clean Ir(111) substrate at a low temperature (~30 K) in which the diffusion of the Fe adatoms is suppressed as compared to the situation at room temperature. Accordingly, the Fe islands exhibit a fractal shape as shown in Fig. 5(a). Although the substrate has the same low impurity as that in Fig. 1(e), the suppressed diffusion of Fe adatoms prevents the formation of triangular-shaped islands. Consequently, no skyrmion lattice is observed in the fractal-shaped Fe islands. This confirms that the diffusivity of the Fe adatoms plays the most critical role in skyrmion formation.

## 4. Conclusions

In summary, we have conducted a systematic study on the influence of diffusivity of Fe adatoms on the formation of skyrmion lattice in Fe islands. We demonstrate that a high substrate surface impurity or a low growth temperature reduces the diffusivity of Fe adatoms yielding islands with either irregular or fractal shapes. The reduced diffusivity will likely affect the perfection of the atomic structure of the Fe islands, which in turn changes the competing magnetic interactions. As a result, the formation condition of the skyrmion lattice is broken and no skyrmion lattice can be observed. In contrast, for fully developed triangular-shaped islands, the adequate diffusion of the Fe adatoms ensures the perfection of the atomic structure and thus the skyrmion lattice formation. These results reveal an intimate interplay between structural and magnetic degrees of freedom in the Fe/Ir(111) system.

**Declaration of competing interest**

The authors declare no competing interests.

**Data availability**

All raw and derived data used to support the findings of this work are available from the authors on reasonable request.

**Acknowledgement**

This work was supported by National Key Research Program of China (2022YFA1403300), the National Natural Science Foundation of China (11427902, 11991060, 12074075, 12074080 and 12274088), the Shanghai Municipal Science and Technology Major Project (2019SHZDZX01), and the Shanghai Municipal Natural Science Foundation (20501130600, 22ZR1408100 and 22ZR1407400)

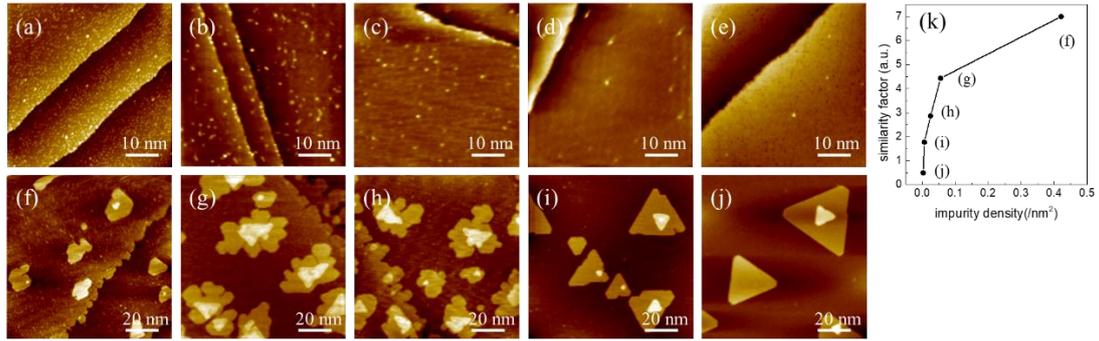

**Fig. 1.** Large scale topography of Fe/Ir(111) system. (a-e) Topography of bare Ir(111) substrate with different surface impurity density. The estimated density level and tunnel parameters are (a) 0.4212/nm$^2$, -1 V, 100 pA, (b) 0.0552/nm$^2$, 0.76 V, 100 pA, (c) 0.0248/nm$^2$, 1 V, 100 pA, (d) 0.006/nm$^2$, 1 V, 100 pA, (e) 0.0024/nm$^2$, 1 V, 100 pA. (f-j) Topography of monolayer Fe island grown on Ir(111) substrate with the surface impurity density displayed on the top. The tunnel parameters are (f) 1V, 100 pA, (g) 1 V, 100 pA, (h) 1 V, 100 pA, (i) 1 V, 100 pA, (j) 0.1 V, 100 pA. (k) The plot of the similarity factor of the Fe monolayer island as a function of the surface impurity density.

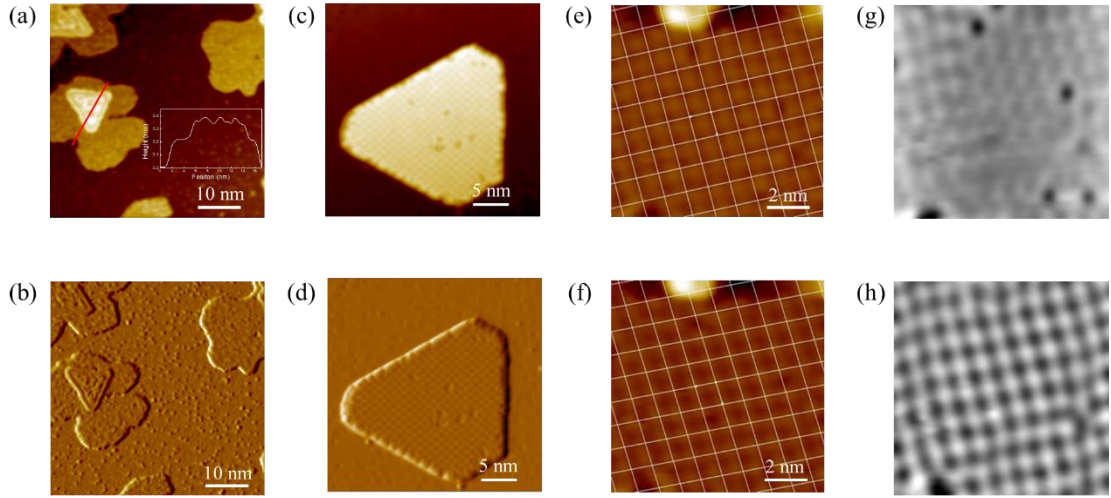

**Fig. 2.** The topography and the corresponding height differential image of Fe monolayer islands. (a-b) topography and its differential of Fe monolayer island grown under a high surface impurity density. A line-cut height profile is shown in the inset. The tunnel parameters are 0.1 V, 100 pA. (c-d) topography and its differential of Fe monolayer island grown under a clean limit. The tunnel parameters are 0.1 V, 100 pA. (e-f) STM constant-current topographic images of the square lattice under +0.25 T and -0.25 T magnetic field, respectively. The tunnel parameters are 0.05 V, 100 pA for these images. (g-h) The sum and the difference of (e) and (f), respectively.

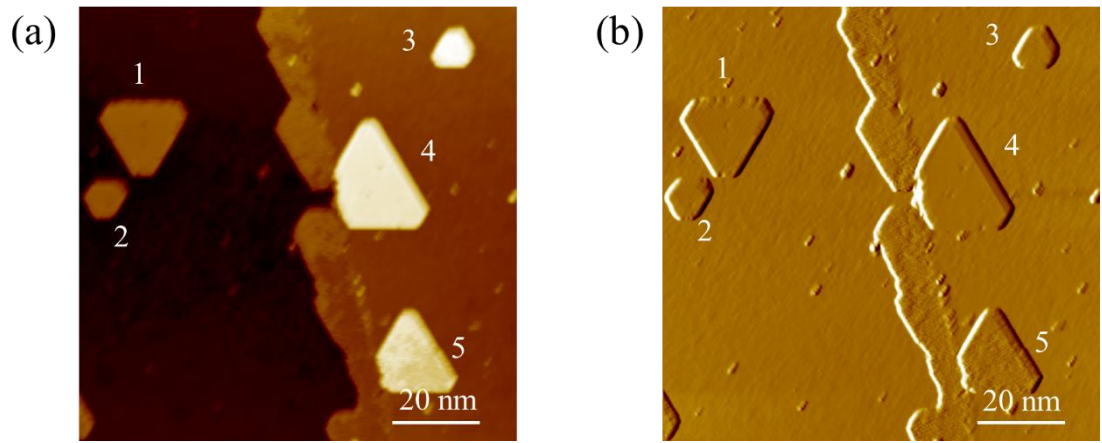

**Fig. 3.** The topography and the corresponding height differential image of Fe monolayer islands grown under an intermediated surface impurity density level. The tunnel parameters are 0.1 V, 100 pA.

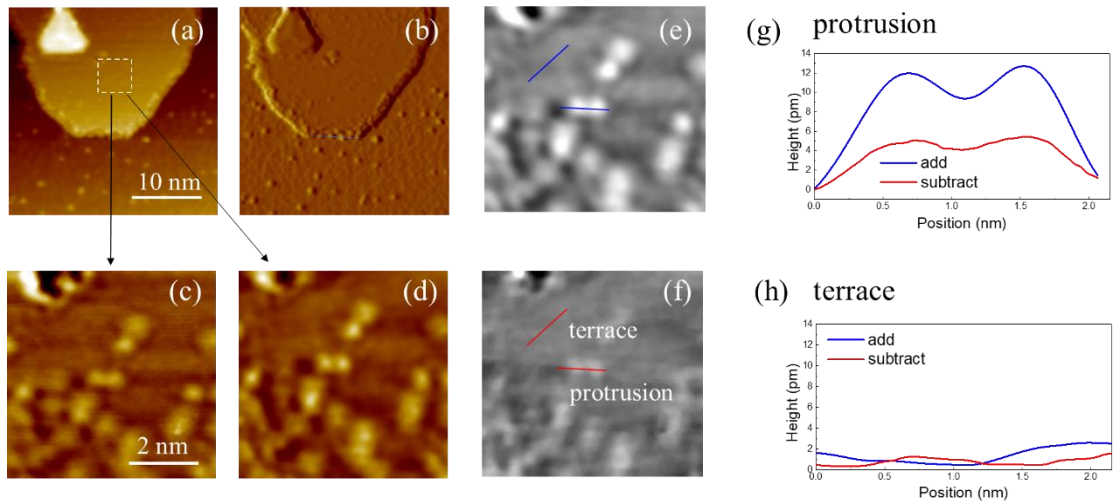

**Fig. 4.** A magnetic field dependent STM measurement on a non-skyrmion phase. (a-b) STM constant-current topographic image and its differential of an Fe island with a non-skyrmion phase. The tunnel parameters are 0.05 V, 100 pA for these images. (c-d) The zoomed-in STM images of the square area denoted in Fig. 4(a). They are taken under -0.25 T and +0.25 T magnetic field, respectively. The tunnel parameters are 0.05 V, 100 pA for these images. (e-f) The sum and the difference of (c) and (d), respectively. (g-h) The line-cut profiles from (e) and (f).

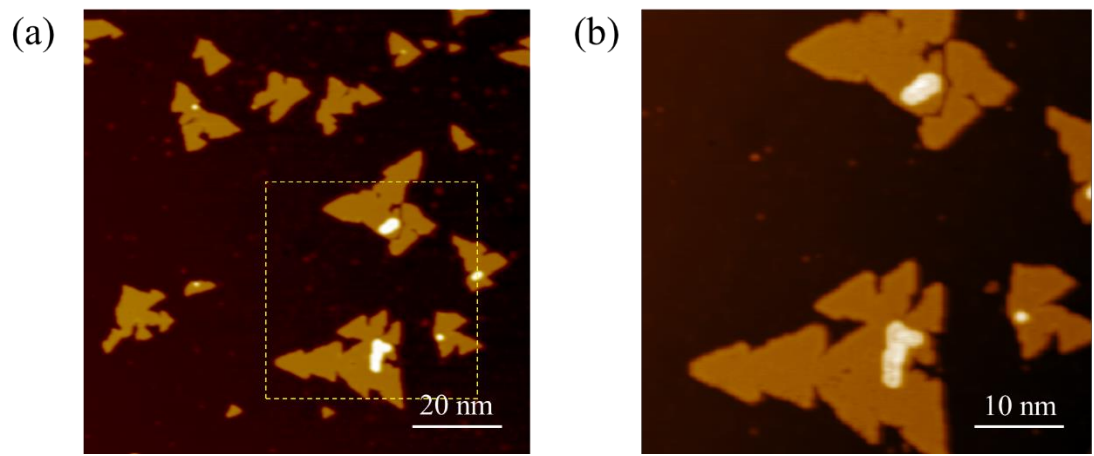

**Fig. 5.** The topography of Fe monolayer islands grown at low temperatures with a clean substrate surface. (b) is a zoomed-in image of the denoted yellow square in (a). The tunnel parameters are 0.05 V, 200 pA for these images.